\title{ic1396_proper_motions}
\documentclass{aa}
\usepackage{babel}
\usepackage[varg]{txfonts}
\usepackage{graphicx}
\usepackage{color}
\newcommand{\RA}[4] {$\alpha(J2000)=#1^\mathrm{h}#2^\mathrm{m}#3\fs#4$}
\newcommand{\DEC}[4]{$\delta(J2000)=#1\degr#2'#3\farcs#4$}

\newcommand{\Msol}  {$M_{\sun}$}

\newcommand{\phn}   {\ensuremath{\phantom{0}}}

\newcommand{\hmol}  {H$_2~\upsilon=1$--0 S(1)}
\newcommand{\hdos}  {H$_2$}
\newcommand{\vlsr}  {\ensuremath{V_\mathrm{LSR}}}

\newcommand{\kms}   {km~s$^{-1}$}

\newcommand{\fdeg}  {\mbox{\ensuremath{.\!\!^\circ}}}

\begin{document}

\title{
Collision of protostellar jets in the star-forming region IC 1396N}
\subtitle{
Analysis of knot proper motions\thanks{
Partially based on service observations (SW2015a35) made with the WHT telescope operated on the island of La Palma by the ING in the ORM of the IAC} 
}

\titlerunning{Collision of protostellar jets in IC 1396N}

\author{
Rosario L\'opez\inst{1}
\and
Robert Estalella\inst{1}
\and
Maria T. Beltr\'an\inst{2} 
\and
Fabricio Massi\inst{2}
\and
Jos\'e A. Acosta-Pulido\inst{3}
\and
Josep M. Girart\inst{4,5}
}

\institute{
Departament de F\'{\i}sica Qu\`antica i Astrof\'{\i}sica,
Institut de Ci\`encies del Cosmos, Universitat de Barcelona, IEEC-UB,
Mart\'{i} i Franqu\`es, 1, E-08028 Barcelona, Spain.
\and
INAF -- Osservatorio Astrofisico di Arcetri, Largo E. Fermi 5, 50125, Firenze,
Italy.
\and
Instituto de Astrof\'{\i}sica de Canarias, 
Avenida V\'{\i}a L\'actea, 38205 La Laguna,  Tenerife, Spain.
\and
Institut de Ci\`encies de l'Espai (ICE), CSIC,
Carrer de Can Magrans, s/n, E-08193, Cerdanyola del Vall\`es, Catalonia, Spain.
\and
Institut d'Estudis Espacials de Catalunya (IEEC), E-08034, Barcelona, Catalonia, Spain.
}

\authorrunning{L\'opez et al.}

\date{
Received 21 February 2022/
Accepted 22 March 2022
}

\abstract{
The bright-rimmed cloud IC 1396N 
is believed to host one of the few known cases where two bipolar CO outflows  driven by young stellar objects actually collide. 
The CO outflows are traced by chains of knots of \hdos{} emission, with enhanced emission at the position of the possible collision.
}{
The aim of this work  is to use the proper motions of the \hdos{} knots to confirm the collision scenario.
}{
A second epoch \hdos{} image was obtained, and the proper motions of the knots
were determined with a time baseline of $\sim11$ years.
We also performed differential photometry on the images to check the flux variability of the knots. 
}{
For each outflow (N and S) we classified the knots as pre-collision or post-collision.
The axes of the pre-collision knots, the position of the possible collision point, and the axes of the post-collision knots were estimated. 
The difference between the proper motion direction of the post-collision knots and the position angle from the collision point was also calculated.
For some of the knots we obtained the 3D velocity by using the radial velocity derived from \hdos{} spectra.
}{
The velocity pattern of the \hdos{} knots in the area of interaction (post-collision 
knots) shows a deviation from that of the pre-collision knots, 
consistent with being a consequence of the interaction between the two outflows. 
This favours the interpretation of the IC 1396N outflows as a true collision between two protostellar jets instead of a projection effect.
}{}


\maketitle

\section{Introduction}

Jets and outflows are ubiquitous in star-forming regions, but the cases of interaction between outflows are extremely rare. In fact, only two cases of apparent collision between jets or outflows are known, the IC 1396N outflows \citep{Bel12} and BHR 71 \citep{Zap18}.
In BHR 71, the red lobe and the blue lobe of two different CO outflows collide, and in the area of interaction there is an increase  of CO emission, and broadening of the lines, consistent with the collision scenario.

IC~1396N is a bright-rimmed cloud \citep[BRC38;][]{Sug91}, located in the Cep OB2 association. 
We adopt an updated value for the distance to IC 1396N of $910\pm49$~pc, obtained from the catalogue of distances to molecular clouds of  \citet{Zuc20}. 
IC 1396N shows a cometary structure elongated in the south-north direction, and is associated with an intermediate-mass star-forming region where a number of Herbig-Haro objects, \hdos{} jet-like features, CO molecular outflows and millimeter compact sources have been reported \citep[see][for a thorough description of the region]{Bel09}.
Deep near-infrared images through the 2.12 $\mu$m \hmol{} line of
IC~1396N carried out with the TNG telescope by \citet{Bel09} reveal a
number of small-scale \hdos{} emission features spread all over the globule,
which are resolved into several chains of knots, showing a jet-like morphology
and possibly tracing different \hdos{} outflows.

\begin{figure}[htb]
\caption{\label{fig_co}
CO outflows (red and blue contours), dust continuum emission (green contours), 
and \hdos{} emission (gray scale) from \citet{Bel12}. 
Source C traces the driving source of outflow N, while 
source I traces the driving source of outflow S.
}
\resizebox{0.99\hsize}{!}{\includegraphics{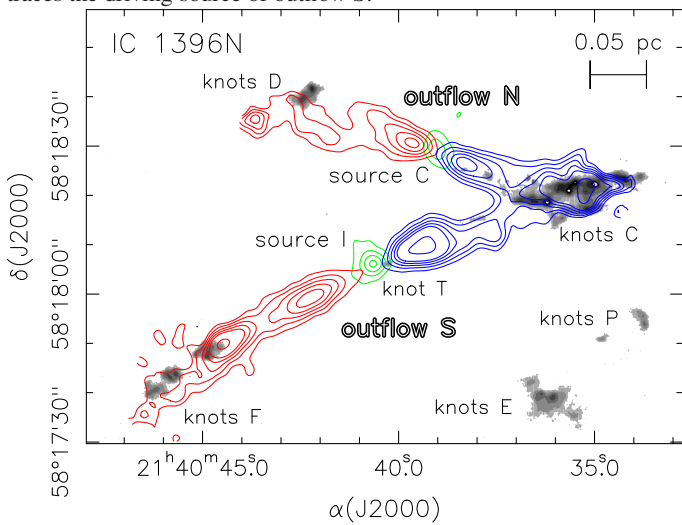}}
\end{figure}

Two well-collimated bipolar CO outflows (outflows N and S) toward the northern part of the globule are reported in \citet{Bel12}. 
In both outflows the blueshifted and redshifted lobes emanate from sources traced by their dust continuum emission (source C for outflow N and source I for outflow S), located at their symmetry center (see Fig.\ \ref{fig_co}).
Both CO outflows are associated with several chains of \hdos{}
knots reported by \citet{Bel09}: the \hdos{}
chains of knots D and F lie along the redshifted lobes of the N and
S outflows respectively, while the chain of knots C lie at the position 
where the CO-North and CO-South blueshifted outflows lobes overlap.

\citet{Bel09,Bel12} propose that the outflows N and S  are colliding at the position of knots C, explaining the strong \hdos{} emission of this chain of knots, although this is not the only possible explanation, since it could be a projection effect, with the two outflows being in different planes.
However, the fact that the two outflows are almost in the plane of the sky (inclination $i<10\degr$), and that the powering sources are at the same systemic velocity ($\vlsr\simeq0$ \kms), favour the collision scenario.

With the aim of shedding new light on whether the overlapping blueshifted lobes are only a projection effect or, on the contrary, are tracing the collision of the two outflows, we decided to obtain the proper motions of the  \hdos{} knots of IC 1396N northern outflows.
Currently, proper motions of jet knots are used as an indirect tool for identifying the outflow driving source in regions with complex outflow activity:
assuming that the knot motions are ballistic, the directions of their proper motion diverge from the location of the driving source. 
In the case of the IC 1396N northern outflows, if the superposition is only a projection effect, we would expect that the pattern of tangential velocities be the same for knots C as for the rest of the knots. 
On the other hand, if there is a true collision, we would expect that the pattern of tangential velocities of knots C to be modified by the collision. 
In order to derive the proper motions of the \hdos{} knots projected onto the outflow lobes we obtained a second-epoch, deep \hmol{} image of IC 1396N. 

In addition, for some of the knots we were able to complement their tangential velocity with the line-of-sight velocity obtained from the spectroscopic \hmol{} observations of \citet{Mas22}.

In \S2 we describe the observations and the derivation of the proper motions,
in \S3 we present the results obtained: we analyze the pattern of proper motions obtained, and 
in \S4 we discuss the results and we give our conclusions.

\section{Observations and Data Reduction} 

The first epoch \hdos{} image of IC 1396N was obtained
with NICS at the 3.58~m Telescopio Nazionale Galileo (TNG) during the observing run of the 2005 July 16 \citep[all the details are given in][]{Bel09}. 
The second epoch \hdos{} image was obtained on 2016 July 14 as a part of the ING service-mode program with LIRIS (\cite{Aco03}; \cite{Man04}) at the  4.2~m Williams Herschel Telescope (WHT) of the Observatorio del Roque de los Muchachos (ORM, La Palma, Spain). 
LIRIS is equipped with a Rockwell Hawaii $1024\times1024$ HgCdTe
array detector. The spatial scale is $0\farcs25$~pixel$^{-1}$, giving an image
field of view (FOV) of $4\farcm27 \times 4\farcm27$. The observing strategy
followed a 3-point dithering pattern on-source interspersed with the same
pattern for an off-source position of a field located $\sim 6$~arcmin from the
target. Three exposures of 50s were taken at each of the dither positions. The
on-off cycle was repeated 6 times to complete a total on-source integration time
of 2700~s. Typical value of the seeing was $\sim$ 0$\farcs$8.

Data were processed using the  package {\it lirisdr} developed by the LIRIS
team within the  {\small IRAF} environment\footnote{{\small 
IRAF} is distributed
by the National Optical Astronomy Observatories, which are operated by the
Association of Universities for Research in Astronomy, Inc., under cooperative
agreement with the National Science Foundation.
}. 
The reduction process included
sky subtraction,  flat-fielding, correction of geometrical distortion, and
finally combination of frames using the common ``shift-and-add''
technique.  This final step consisted in dedithering and co-adding frames
taken at different dither points to obtain a mosaic for each filter.  The
resulting mosaic covered a FOV of $\sim 5$ arcmin$^2$.

\subsection{Proper motions determination}

Prior to evaluate knot proper motions, the two \hdos{} IC1396 images were
converted into a common reference system. 
The positions of  twelve field stars, common to  the two frames, were used to register the images. 
The {\it geomap} and  {\it geotran}  tasks of {\small IRAF} were applied to perform a linear transformation, with six free parameters that take into account translation, rotation and magnification between different frames. 
After the transformation, the typical rms of the difference in position for the reference stars between the  two images was $\sim0.08$~arcsec in both coordinates. 
The two images had the same spatial scale ($0.25$ arcsec~pixel$^{-1}$), which was preserved by the transformation.

We defined boxes enclosing the emission of each of the \hdos{} knots that were
identified in \citet{Bel09}. Then, the position offset in the $x$ and $y$ coordinates
between the two epochs was calculated by cross-correlation.  The uncertainty in
the position of the correlation peak was estimated  through the scatter of the
correlation peak positions obtained from boxes differing from the nominal one by
$\pm2$ pixels. The error adopted for each coordinate for the offset between the
two epochs was twice the uncertainty in the correlation peak position, added
quadratically to the rms alignment error  
\citep[see the description of the used method in][]{Ang07}.

\subsection{Differential photometry}

We carried out photometry on the \hdos{} images to check the variability of the line emission from the knots.
In addition to our images, we also used an \hdos{} image, obtained with NICS at the TNG on 2003 October 17 \citep{Car06}, kindly provided by the authors. 
Continuum was not subtracted as the knots are barely affected by continuum emission, in order not to add additional errors. 
The three images were registered to a common reference frame as described in \S2.1. 
Then, we performed aperture photometry on a sample of $\sim200$ common stars using the {\it phot} task of {\small IRAF}. 
The zero points were obtained as a weighted (SNR) mean of the instrumental magnitudes, after excluding stars deviating more than  $5\sigma$ to minimise the effects of variable stars. 
As we were only interested in differential photometry, we did not calibrate the zero points \citep[the absolute knot photometry is already given in][]{Bel09}.

The photometry of the knots was performed using the {\it polyphot} task of {\small IRAF}. 
For each knot, we defined a polygon on the 2005 image enclosing the emission down to a  $5\sigma$ level and used the same polygons and sky annuli throughout the three images, as the image are on the same reference frame. 
We also checked that the polygons were large enough to be not affected by knots proper motions. The results are listed in Table \ref{tab_phot} of the Appendix.

\section{Results}

\subsection{Proper motions}

Proper motions were determined for all the knots detected in the two images, except for knot C6, which changed its morphology between the two epochs (see Fig.\ \ref{fig_c6}). The spurious proper motion determined for this knot will not be considered for the analysis.

\begin{table*}[htb]
\small
\centering
\caption{\label{tab_pm}\label{tab_vtot}
Proper motions measured in IC 1396, tangential velocities for a distance of 910 pc, and 3D velocities for the knots whose radial velocity is known.
}
\addtolength{\tabcolsep}{-1pt}
\begin{tabular}{llr@{$\,\pm\,$}lr@{$\,\pm\,$}lr@{$\,\pm\,$}lr@{$\,\pm\,$}lr@{$\,\pm\,$}lr@{$\,\pm\,$}lr@{$\,\pm\,$}lr@{$\,\pm\,$}lr@{$\,\pm\,$}l}
\hline
\hline
& &
\multicolumn{2}{c}{$\mu_x$} &
\multicolumn{2}{c}{$\mu_y$} &
\multicolumn{2}{c}{$v_x$} & 
\multicolumn{2}{c}{$v_y$} &
\multicolumn{2}{c}{$v_t$\tablefootmark{a}} & 
\multicolumn{2}{c}{PA\tablefootmark{a}}    & 
\multicolumn{2}{c}{$v_r$\tablefootmark{b}} & 
\multicolumn{2}{c}{$v_\mathrm{tot}$\tablefootmark{c}}&
\multicolumn{2}{c}{$i$\tablefootmark{c}}\\
Knot  & Id.\tablefootmark{d} &
\multicolumn{2}{c}{(mas yr$^{-1}$)} & 
\multicolumn{2}{c}{(mas yr$^{-1}$)} & 
\multicolumn{2}{c}{(\kms)} & 
\multicolumn{2}{c}{(\kms)} & 
\multicolumn{2}{c}{(\kms)} & 
\multicolumn{2}{c}{(deg)}  & 
\multicolumn{2}{c}{(\kms)} & 
\multicolumn{2}{c}{(\kms)} &
\multicolumn{2}{c}{(deg)}\\
\hline
D1	&	N-red	& $-4.23$&$1.65$ & $ 3.46$&$1.25$ & $-18.2$&$ 7.1$ & $ 14.9$&$ 5.4$  & $23.6$&$ 6.5$ & $  51$&$14$ & \multicolumn{2}{c}{$\cdots$} & \multicolumn{2}{c}{$\cdots$} & \multicolumn{2}{c}{$\cdots$} \\
D2	&	N-red	& $-2.64$&$1.42$ & $ 0.90$&$0.68$ & $-11.4$&$ 6.1$ & $  3.9$&$ 2.9$  & $12.1$&$ 5.9$ & $  71$&$14$ & \multicolumn{2}{c}{$\cdots$} & \multicolumn{2}{c}{$\cdots$} & \multicolumn{2}{c}{$\cdots$} \\
C1	&	N-blue	& $ 8.68$&$9.30$ & $ 2.90$&$2.54$ & $ 37.5$&$40.1$ & $ 12.5$&$10.9$  & $39.5$&$38.2$ & $ -72$&$24$ & \multicolumn{2}{c}{$\cdots$} & \multicolumn{2}{c}{$\cdots$} & \multicolumn{2}{c}{$\cdots$} \\
C3	&	N-blue	& $11.00$&$4.70$ & $-2.39$&$2.51$ & $ 47.4$&$20.3$ & $-10.3$&$10.8$  & $48.6$&$19.9$ & $-102$&$14$ & \multicolumn{2}{c}{$\cdots$} & \multicolumn{2}{c}{$\cdots$} & \multicolumn{2}{c}{$\cdots$} \\
C4	&	N-blue	& $-3.86$&$3.26$ & $-4.87$&$5.28$ & $-16.6$&$14.1$ & $-21.0$&$22.8$  & $26.8$&$19.9$ & $ 142$&$39$ & \multicolumn{2}{c}{$\cdots$} & \multicolumn{2}{c}{$\cdots$} & \multicolumn{2}{c}{$\cdots$} \\
C5	&	N-blue	& $ 2.27$&$1.06$ & $-2.28$&$1.29$ & $  9.8$&$ 4.6$ & $ -9.8$&$ 5.6$  & $13.9$&$ 5.1$ & $-135$&$23$ & \multicolumn{2}{c}{$\cdots$} & \multicolumn{2}{c}{$\cdots$} & \multicolumn{2}{c}{$\cdots$} \\
\hline
F1	&	S-red	& $-1.26$&$0.92$ & $ 2.02$&$0.70$ & $ -5.4$&$ 4.0$ & $  8.7$&$ 3.0$  & $10.3$&$ 3.3$ & $  32$&$20$ & $  9$&$28$                   & $  14$&$19$                  & $ 41$&$89$                   \\
F2-3 &	S-red	& $-2.73$&$1.09$ & $-5.47$&$1.22$ & $-11.8$&$ 4.7$ & $-23.6$&$ 5.3$  & $26.4$&$ 5.2$ & $ 154$&$11$ & $-11$&$28$                   & $  29$&$12$                  & $-23$&$52$                   \\
F4	&	S-red	& $-4.95$&$2.87$ & $ 0.84$&$1.51$ & $-21.4$&$12.4$ & $  3.6$&$ 6.5$  & $21.7$&$12.2$ & $  80$&$17$ & $-33$&$28$                   & $  39$&$24$                  & $-57$&$27$                   \\
F5	&	S-red	& $-4.78$&$3.26$ & $ 0.64$&$1.23$ & $-20.6$&$14.1$ & $  2.8$&$ 5.3$  & $20.8$&$14.0$ & $  82$&$14$ & $  4$&$28$                   & $  21$&$15$                  & $ 11$&$75$                   \\
T	&	S-blue	& $	6.15$&$1.32$ & $ 2.46$&$0.90$ & $ 26.5$&$	5.7$ & $ 10.$& $ 3.9$  & $28.6$&$ 5.5$ & $ -68$&$ 8$ & \multicolumn{2}{c}{$\cdots$} & \multicolumn{2}{c}{$\cdots$} & \multicolumn{2}{c}{$\cdots$} \\
C2	&	S-blue	& $ 9.66$&$3.39$ & $ 6.92$&$2.86$ & $ 41.7$&$14.6$ & $ 29.8$&$12.4$  & $51.3$&$13.9$ & $ -54$&$15$ & \multicolumn{2}{c}{$\cdots$} & \multicolumn{2}{c}{$\cdots$} & \multicolumn{2}{c}{$\cdots$} \\
\hline
C6	&	N-post	& $-4.29$&$2.68$ & $ 3.83$&$1.94$ & $-18.5$&$11.6$ & $ 16.5$&$ 8.4$  & $24.8$&$10.3$ & $  48$&$24$ & $-47$&$23$                   & $  53$&$21$                  & $-62$&$15$                   \\
C7	&	N-post	& $ 7.06$&$2.52$ & $-3.28$&$0.95$ & $ 30.4$&$10.9$ & $-14.1$&$ 4.1$  & $33.6$&$10.0$ & $-115$&$10$ & \multicolumn{2}{c}{$\cdots$} & \multicolumn{2}{c}{$\cdots$} & \multicolumn{2}{c}{$\cdots$} \\
C9	&	N-post	& $-1.18$&$1.43$ & $ 0.49$&$0.86$ & $ -5.1$&$ 6.2$ & $  2.1$&$ 3.7$  & $ 5.5$&$ 5.9$ & $  67$&$43$ & \multicolumn{2}{c}{$\cdots$} & \multicolumn{2}{c}{$\cdots$} & \multicolumn{2}{c}{$\cdots$} \\
\hline
C8	&	S-post	& $ 2.15$&$1.58$ & $ 1.91$&$1.26$ & $  9.3$&$ 6.8$ & $  8.2$&$ 5.4$  & $12.4$&$ 6.3$ & $ -48$&$28$ & $-25 $&$23$                  & $  28$&$21$                  & $-64$&$24$                   \\
C10	&	S-post	& $ 6.45$&$2.37$ & $ 1.68$&$0.90$ & $ 27.8$&$10.2$ & $  7.3$&$ 3.9$  & $28.7$&$10.0$ & $ -75$&$ 9$ & $-43 $&$23$                  & $  52$&$20$                  & $-56$&$17$                   \\
C11	&	S-post	& $ 1.39$&$1.25$ & $-1.22$&$1.03$ & $  6.0$&$ 5.4$ & $ -5.3$&$ 4.4$  & $ 8.0$&$ 5.0$ & $-132$&$35$ & \multicolumn{2}{c}{$\cdots$} & \multicolumn{2}{c}{$\cdots$} & \multicolumn{2}{c}{$\cdots$} \\
C12	&	S-post	& $ 0.83$&$0.76$ & $ 0.83$&$0.96$ & $  3.6$&$ 3.3$ & $  3.6$&$ 4.1$  & $ 5.1$&$ 3.7$ & $ -45$&$41$ & \multicolumn{2}{c}{$\cdots$} & \multicolumn{2}{c}{$\cdots$} & \multicolumn{2}{c}{$\cdots$} \\
C13	&	S-post	& $	2.02$&$0.80$ & $ 1.84$&$1.10$ & $  8.7$&$ 3.4$ & $  8.0$&$ 4.7$  & $11.8$&$ 4.1$ & $ -48$&$20$ & \multicolumn{2}{c}{$\cdots$} & \multicolumn{2}{c}{$\cdots$} & \multicolumn{2}{c}{$\cdots$} \\
C14 &	S-post	& $ 8.11$&$2.07$ & $-0.41$&$0.75$ & $ 35.0$&$ 9.0$ & $ -1.8$&$ 3.3$  & $35.0$&$ 8.9$ & $ -93$&$ 6$ & $ -8 $&$23$                  & $  36$&$10$                  & $-13$&$36$                   \\
C15	&	S-post	& $ 3.77$&$3.43$ & $ 1.40$&$1.32$ & $ 16.2$&$14.8$ & $  6.0$&$ 5.7$  & $17.3$&$14.0$ & $ -70$&$25$ & \multicolumn{2}{c}{$\cdots$} & \multicolumn{2}{c}{$\cdots$} & \multicolumn{2}{c}{$\cdots$} \\
C16	&	S-post	& $ 2.86$&$0.65$ & $ 1.98$&$0.63$ & $ 12.3$&$ 2.8$ & $  8.5$&$ 2.7$  & $15.0$&$ 2.8$ & $ -55$&$10$ & $-10 $&$23$                  & $  18$&$13$                  & $-35$&$62$                   \\
C17	&	S-post	& $ 5.39$&$2.51$ & $ 1.62$&$1.05$ & $ 23.3$&$10.8$ & $  7.0$&$ 4.5$  & $24.3$&$10.4$ & $ -73$&$13$ & \multicolumn{2}{c}{$\cdots$} & \multicolumn{2}{c}{$\cdots$} & \multicolumn{2}{c}{$\cdots$} \\
\hline
\end{tabular}
\addtolength{\tabcolsep}{1pt}
\tablefoot{ 
\tablefoottext{a}
{Tangential velocity and position angle, eastward from North.}
\tablefoottext{b}
{Radial velocity from \citet{Mas22}; the errors quoted are the random plus systematic errors.}
\tablefoottext{c}
{3D velocity; $i$ is the inclination angle with respect to the plane of the sky, positive away from the observer.}
\tablefoottext{d}
{Identification of the knot: outflow (N or S), and lobe, red, blue, or post-collision.}
}
\end{table*}

The proper motions measured are shown in Table \ref{tab_pm} and  Figs.\ \ref{fig_knots} and \ref{fig_zoom_knots}. 
In order to analyze the proper motions, we will segregate the knots in different groups: 
outflow N pre-collision knots, from the east (redshifted) and west (blueshifted) lobes (N-red, N-blue), 
outflow S pre-collision knots, from the east (redshifted) and west (blueshifted) lobes (S-red, S-blue), and 
post-collision knots in the interaction area. 
The post-collision knots are tentatively identified as belonging 
to outflow N (N-post: knots C6, C7, C9) or 
to outflow S (S-post: C8 and C10 to C17). 
This identification will be justified later on, and is shown in Table \ref{tab_pm}.

\subsection{Direction of the outflows axes}

\subsubsection{Pre-collision knots}

\begin{table*}[htb]
\centering
\caption{\label{tab_preaxis}
Pre-collision outflows axes, passing through the position of each exciting source.
}
\begin{tabular}{lcclr@{$\,\pm\,$}lcll}
\hline
\hline
& Mean ${v_t}$\tablefootmark{a} 
& Driving 
&      
& \multicolumn{2}{c}{Axis PA\tablefootmark{b}} 
& Bending\tablefootmark{c} & 
${\langle d^2\rangle^{1/2}}$\tablefootmark{(d)} &\\
Outflow& (\kms) & source & Lobe & \multicolumn{2}{c}{(deg)} & (deg) & (arcsec) & Knots used \\
\hline
N  & $19.2\pm4.1$ & C  & Red  & $  67.6$&$0.3$ &              & 1.3 & D1, D2\\
   &              &    & Blue & $-122.3$&$0.5$ & $9.8\pm0.6$  & 0.5 & C1, C3, C4, C5\\
\hline
S  & $20.5\pm2.2$ & I  & Red  & $ 118.9$&$0.2$ &              & 0.9 & F1, F2-3, F4, F5\\
   &              &    & Blue & $ -66.7$&$0.3$ & $5.6\pm0.3$  & 0.7 & T, C2\\            
\hline
\end{tabular}
\tablefoot{ 
\tablefoottext{a}
{Scalar average of tangential velocities $v_t$.}
\tablefoottext{b}
{PA of the best-fit half-line with origin at the corresponding exciting source.}
\tablefoottext{c}
{Difference in direction between the blue and red lobes.}
\tablefoottext{d}
{Root-mean-square distance of the knots to the axis (see Table \ref{tab_distpre}).}
}
\end{table*}

Following the nomenclature of \citet{Bel12}, outflow N is powered by source C (traced by dust continuum emission) with coordinates 
\RA{21}{40}{39}{04},
\DEC{+58}{18}{29}{8}.
The knots D1, D2 are moving eastward (CO red lobe), and  C1, C3, C4, C5, are moving westward (CO blue lobe), toward the area of interaction (see Fig.\ \ref{fig_co}). 
We determined the direction of the E and W lobes separately, allowing for a bending of the outflow at the position of the driving source C. 
We fitted two half lines with origin at the position of the driving source, and minimizing the sum of the squares of the distances to the half line of the positions of knots D1, D2, and of the positions of C1, C3, C4, C5. 
The results are shown in Table \ref{tab_preaxis}.
As can be seen, the axes of the two lobes of outflow N have PA that do not differ exactly at  180 degrees, but they are bent with an angle $9\fdeg8\pm0\fdeg6$ at the position of the driving source C.

The powering source of outflow S is source I, with coordinates
\RA{21}{40}{40}{66}, 
\DEC{+58}{18}{06}{0}. 
The knots F1, F2-F3, F4, F5 are moving eastward (CO red lobe), and  T and C2 westward (CO blue lobe), toward the area of interaction (see Fig.\ \ref{fig_co}). 

We fitted two half lines with origin at the position of the source I
to the positions of knots F1, F2-F3, F4, F5, and of knots T and C2, obtaining the results shown in Table \ref{tab_preaxis}. 
As can be seen, the axes of the two lobes of outflow S are bent with an angle $5\fdeg6\pm0\fdeg3$ at the position of the driving source I.
Both outflows N and S are bent at their driving source with a concavity pointing roughly southward.

The outflow axes are also traced by the high-velocity CO line observed by \citet{Bel12}. 
However, the CO is extended, the axes of the lobes can only be determined reliably where it is more collimated, close to the exciting sources,
and the uncertainty in the axes direction from the CO is higher than that obtained from \hdos{} knots positions.
Thus, we used the latter to estimate collision point (see next section).

\subsubsection{Collision point}

\begin{figure}[htb]
\caption{\label{fig_c6}
\hdos{} emission of knots C5, C6, and C7 in the two epochs. 
The angular scale is indicated in the top panel. 
The morphology of knot C6 changes noticeably between the two epochs.
}

\resizebox{0.99\hsize}{!}{\includegraphics{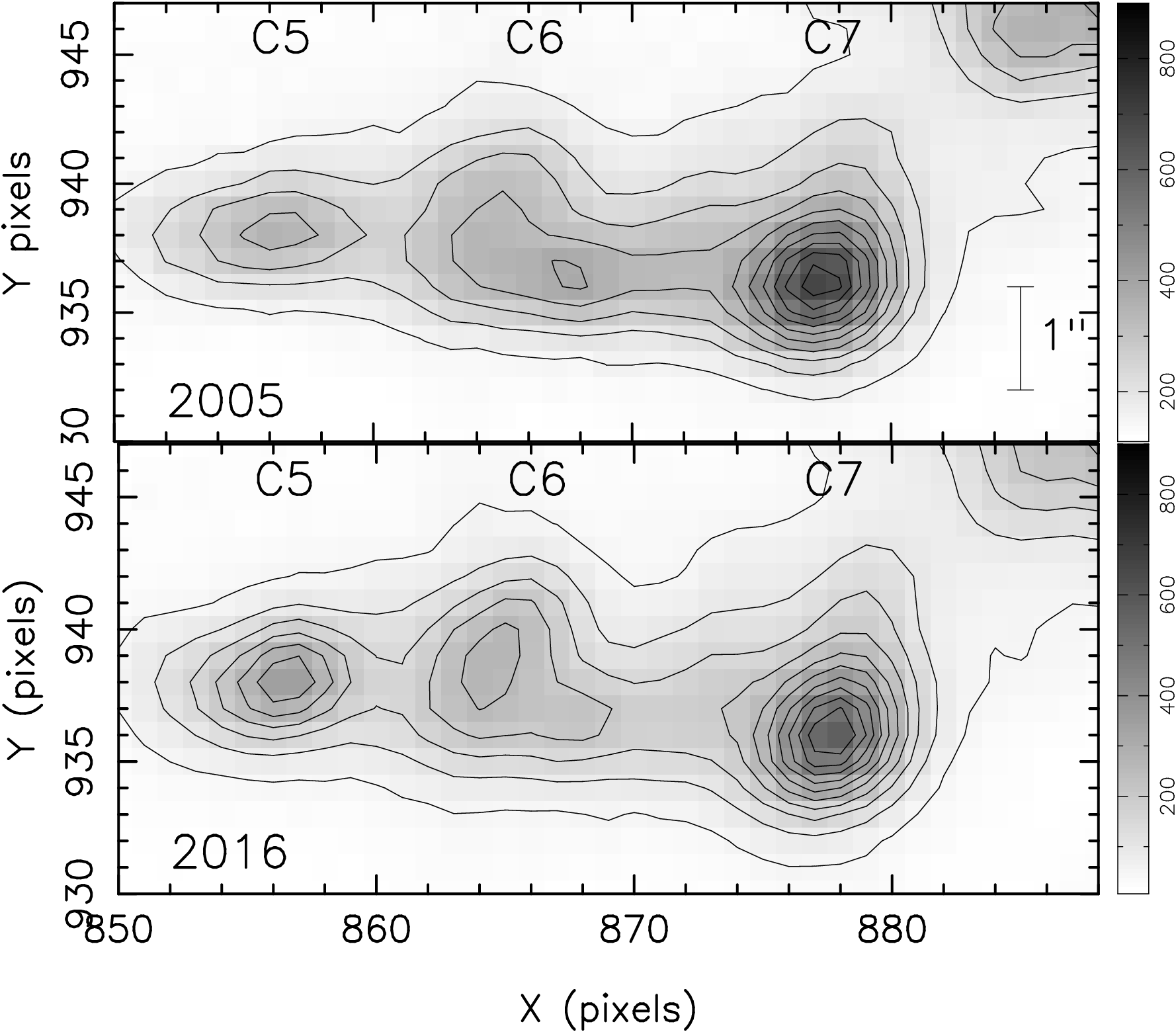}}
\end{figure}

The (possible) collision point of the two outflows  lies at the intersection of the N-blue and S-blue lobes. 
The direction of the S-blue axis is determined from the position of the driving source and only two knots, making its value less reliable that that of he N-blue axis.
Depending on the direction of S-blue axis adopted, the collision point nearly coincides with knot C5,  or is $\sim 1''$ south of knot C6.

Interestingly, both knots show evidence of the interaction of the two outflows. 
On the one hand, the photometric data show that C5 is the only knot that shows variability between the three epochs, and is the knot with the highest increase in brightness (see Table \ref{tab_phot}).
On the other hand, knot C6 is the only knot that changes dramatically its morphology between the two epochs observed, as can be seen in Fig.\ \ref{fig_c6}.
This suggests that these changes are possibly a consequence of the interaction of the two outflows, and most probably the closest knots to the collision point.
The size of the collision region, taken as the separation of knots C5 and C6, is of the order of 5000 au.
We adopted as possible collision point a point of the N-blue axis close to the midpoint between knots C5 and C6. This point has coordinates 
\RA{21}{40}{36}{71}, 
\DEC{+58}{18}{18}{1}. 
By adopting this collision point, knot C5 has to be considered as a pre-collision knot, and knots C6 to C17 as post-collision knots. 
The classification of knot C5 as pre-collision or post-collision is not critical for any of the results of this work.

\begin{table*}[htb]
\centering
\caption{\label{tab_postaxis}
Post-collision outflows axes passing through the collision point,
\RA{21}{40}{36}{71},
\DEC{+58}{18}{18}{1}.
}
\begin{tabular}{ccr@{$\,\pm\,$}cr@{$\,\pm\,$}lcl}
\hline
\hline
& Mean ${v_t}$\tablefootmark{a} 
& \multicolumn{2}{c}{Axis PA\tablefootmark{b}}
& \multicolumn{2}{c}{Deflection\tablefootmark{c}}
& $\langle d^2\rangle^{1/2}$\tablefootmark{(d)} &\\
Outflow& (\kms) & \multicolumn{2}{c}{(deg)} & \multicolumn{2}{c}{(deg)} & (arcsec) & Knots used \\
\hline
N & $12.7\pm5.1$ & $   -106.7$&$0.8$ & $15.6$&$1.0$ & 1.2 & C6\tablefootmark{e}, C7, C9\\
S & $12.9\pm1.7$ & $\phn-74.7$&$0.2$ & $-8.0$&$0.3$ & 1.6 & C8, C10-17\\
\hline
\end{tabular}
\tablefoot{ 
\tablefoottext{a}
{Scalar average of tangential velocity $v_t$.}
\tablefoottext{b}
{PA of the best-fit half-line with origin at the position of the
collision point.}
\tablefoottext{c}
{Difference in position angle between the post-collision axis and the pre-collision axis of the blue lobe from Table \ref{tab_preaxis}.}
\tablefoottext{d}
{Root-mean square distance of the knots to the axis (see Table \ref{tab_distpost}).}
\tablefoottext{e}
{For knot C6, only its position is used.} 
}
\end{table*}

\begin{figure*}[htb]
\caption{\label{fig_knots}
Gray scale: First epoch \hdos{} image of IC 1396N obtained with the TNG.
Plot of the positions and tangential velocity of the knots, derived from their proper motions, for a distance of 910 pc. 
The ellipses at the tip of each arrow indicate the uncertainty in the velocity.
The axes of the red and blue lobes of outflow N are indicated respectively by the orange and magenta dash-dotted lines, passing through the position of its exciting source, C. 
The axes of the red and blue lobes of outflow S are indicated respectively by the red and blue blue dash-dotted lines, passing through the position of its exciting source, I.
The large green stars signs mark the positions of the exciting sources C and I.
The small black or white `+' mark the position of the knots.
An enlarged view of the interaction area can be seen in Fig.\ \ref{fig_zoom_knots}, where the identification of knots C6 to C17 is also shown.
}

\resizebox{\hsize}{!}{\includegraphics{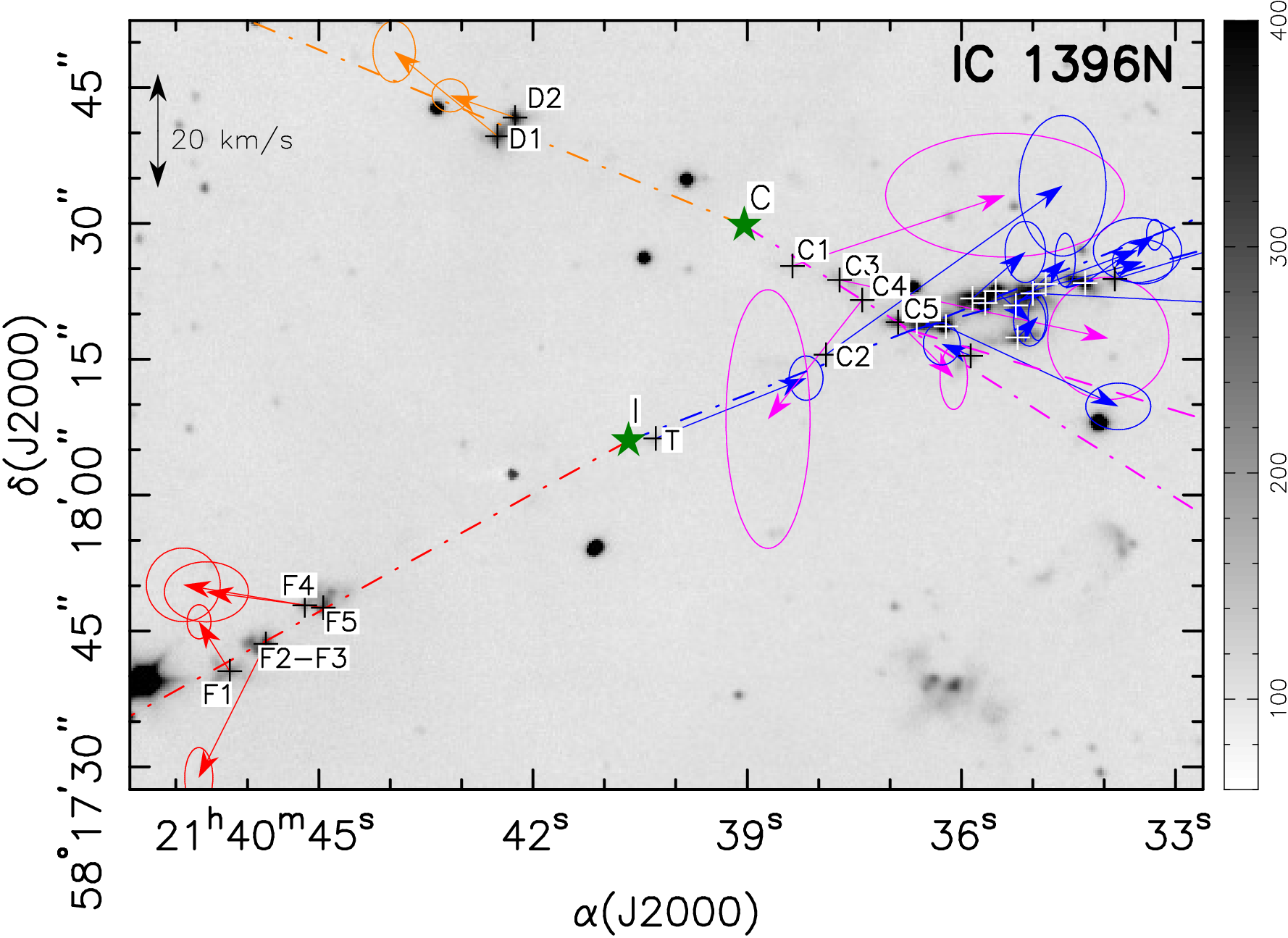}}
\end{figure*}

\begin{figure*}[htb]
\caption{\label{fig_zoom_knots}\label{fig_zoom_axes}
Zoom of Fig.\ \ref{fig_knots} showing the knots in the area of  interaction between the two outflows and their proper motions. 
The blue-lobe pre-collision axes of outflows N and S are indicated respectively by the magenta and blue dash-dotted lines. 
Their intersection, the collision point O, is  marked with a green `*' sign.
The dashed lines show the post-collision axes of outflow N (magenta) and outflow S (blue) (see Table \ref{tab_postaxis}).
}
\resizebox{\hsize}{!}{\includegraphics{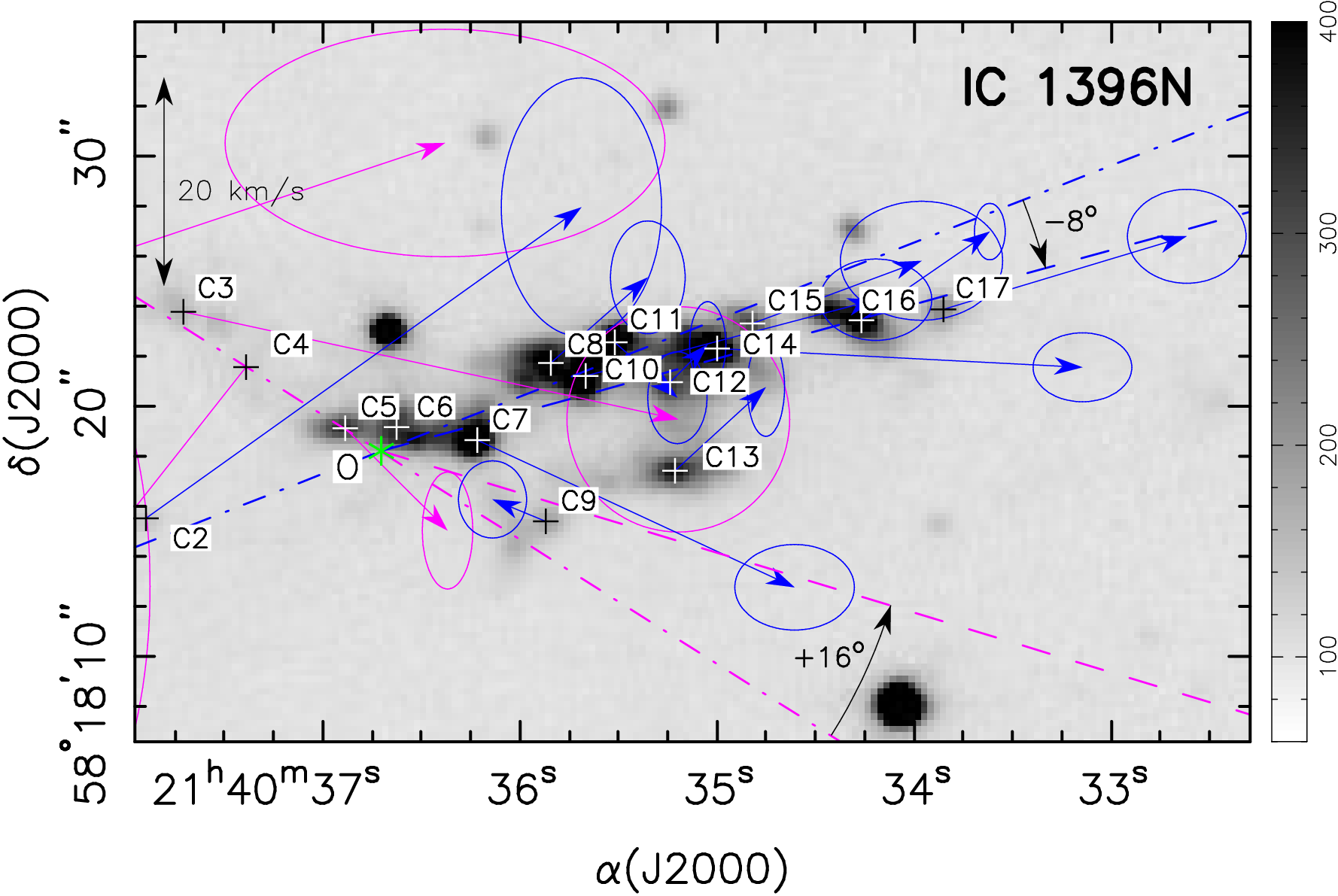}}
\end{figure*}

\subsubsection{Post-collision knots}

As can be seen in Figs.\ \ref{fig_knots} and \ref{fig_zoom_knots}, all the post-collision 
knots are located between the blue lobe axes of
outflow N and outflow S. The knots C6, C7 and C9 are aligned
near the blue lobe axis of outflow N, and because of this we have identified them as
belonging to outflow N.  The rest of post-collision knots, C8 and C10 to C17,
are clearly located to the south of the blue lobe axis of outflow S, but aligned
roughly in the same direction, and because of this we have identified them as
belonging to outflow S. 

In order to characterize the alignment of the post-collision knots, we fitted a
half line, with origin at the position of the collision point, 
minimizing the sum of the squares of the distances to the line of the positions
of the post-collision knots of outflow N, and of the post-collision knots of outflow S.
The results are shown in Table \ref{tab_postaxis}. As can be seen, the
two groups of post-collision knots are well aligned with their axis, with average
distances of $\sim1''$, close to the values for the pre-collision knots (Table
\ref{tab_preaxis}). 

The change in direction of outflow N at the collision is of 
$+15\fdeg6\pm1\fdeg0$ (counterclockwise, northward), while that of outflow S is $-8\fdeg0\pm0\fdeg3$ (clockwise, southward).
This change in direction of the post-collision knots of both outflows is a clear indication that there is an interaction between the two outflows,  giving support to the scenario of collision, where the knots of outflow N are pushed northward, and the knots of outflow S are pushed southward from their original pre-collision direction.

\subsection{Proper motion directions with respect to the outflow axis}

Let us compare the direction of the tangential velocity  of the knots with the directions of the outflow lobes previously derived. For this, we calculated the direction of the average tangential  velocity of the
pre-collision knots of the E lobe of outflows N and S, and
pre-collision and post-collision knots of the W lobe of outflows N and S.
The average was obtained taking into account the error in each value of the tangential velocities. 
The results obtained are shown in Table \ref{tab_pa}.

\begin{table*}[htb]
\centering
\caption{\label{tab_pa}
Vectorial average of the tangential  velocities of the pre-collision and post-collision knots of outflows N and S, and deviation $\langle\Delta\text{PA}\rangle$ from the corresponding outflow-lobe axis.
}
\begin{tabular}{lr@{$\,\pm\,$}lr@{$\,\pm\,$}lr@{$\,\pm\,$}lr@{$\,\pm\,$}lr@{$\,\pm\,$}ll}
\hline
\hline
& 
\multicolumn{2}{c}{Axis PA\tablefootmark{a}} &
\multicolumn{2}{c}{$\langle v_x\rangle$} & 
\multicolumn{2}{c}{$\langle v_y\rangle$} & 
\multicolumn{2}{c}{$\langle\text{PA}\rangle$} &
\multicolumn{2}{c}{$\langle\Delta\mathrm{PA}\rangle$} & \\
Ident.\tablefootmark{b} & 
\multicolumn{2}{c}{(deg)}  & 
\multicolumn{2}{c}{(\kms)} & 
\multicolumn{2}{c}{(\kms)} & 
\multicolumn{2}{c}{(deg)}  & 
\multicolumn{2}{c}{(deg)}  & 
Knots\\
\hline

N-red  & $  67.6$&$0.3$ & $-14.3$&$4.6$ & $ 6.4$&2.6 & $  65.9$& 11.1 & $ -1.7$&11.1 &	D1, D2\\
N-blue & $-122.3$&$0.5$ & $  9.4$&$4.2$ & $-6.7$&4.4 & $-125.6$& 21.7 & $ -3.3$&21.7 & C1, C3, C4, C5\\
S-red  & $ 118.9$&$0.2$ & $ -9.3$&$2.9$ & $ 1.3$&2.2 & $  81.8$& 13.6 & $-37.1$&13.6 & F1, F2-3, F4, F5\\
S-blue & $ -66.7$&$0.3$ & $ 28.5$&$5.3$ & $12.3$&3.7 & $ -66.6$&  7.4 & $  0.1$& 7.4 & T, C2\\
N-post & $-106.7$&$0.8$ & $  3.5$&$5.4$ & $-5.2$&2.8 & $-145.9$& 42.6 & $-39.2$&42.6 & C6, C7, C9\\
S-post & $ -74.7$&$0.8$ & $ 10.1$&$1.6$ & $ 4.6$&1.3 & $ -65.8$&  7.1 & $  8.9$& 7.1 & C8, C10-17\\
\hline
\end{tabular}
\tablefoot{ 
\tablefoottext{a}
{From Table \ref{tab_preaxis}
}
\tablefoottext{b}
{Identification of the knot: outflow (N or S), and lobe, red, blue, or post-collision.}
}
\end{table*}

For the pre-collision knots, we can see that the deviation of the direction of the
tangential velocity with respect to their respective axis has a large
dispersion, but with an average value compatible with zero. 

For the post-collision knots, the uncertainty in their average PA is so large 
that the result is inconclusive, although their average PA is compatible with
that of the pre-collision knots.

\subsection{
Deviation with respect to the radial direction from the collision point}

\begin{table}[htb]
\centering
\caption{\label{tab_rad}
Average deviation of the tangential velocity of the post-collision knots
from the radial direction from the collision point, 
\RA{21}{40}{36}{71},
\DEC{+58}{18}{11}{1} (see Table \ref{tab_radpost}).
}
\begin{tabular}{lrrl}
\hline
\hline
        & Mean $\Delta$PA   & Rms\\
Ident.  & (deg)             & (deg) & Knots\\
\hline
N-post     & $-20.5\pm9.8$ & $48.4$ & C7, C9\\
S-post     & $ -5.8\pm3.8$ & $21.2$ & C8, C10-17\\
\hline
(N+S)-post & $ -7.8\pm3.6$ & $26.4$ & C7-17\\
\hline
\end{tabular}
\end{table}

Let us analyze whether the tangential velocities of the post-collision knots are distributed radially from the collision point. 

Firstly, if we assume that the post-collision knots are ballistic, the quadratic average distance of their trajectories to the collision point should be small. 
Its value is $\langle d^2\rangle^{1/2}= 5\farcs4$ (see Table \ref{tab_postdist}), and thus, the result is not conclusive.

Secondly, we computed the
difference between the tangential velocity position angle and the radial
direction from the collision point. The results are shown in Table
\ref{tab_rad}. 
The average value of the angle difference is small,
$-7\fdeg8\pm3\fdeg6$, but the dispersion is high, $26\degr$. 
Although the dispersion of the direction of the proper motions is high, they are on average radially distributed from the collision point.
Thus, the distribution of the tangential velocities of the post collision knots is consistent with the scenario of a collision of the two outflows at the collision point.

\subsection{Full spatial velocities} 

Table~\ref{tab_vtot} displays the full velocity, $v_\mathrm{tot}$, for the knots for  which the radial velocity, $v_r$, was derived from the \hmol{} spectra of \citet{Mas22}. 
In addition, we derived the inclination angle, $i$, between the knot motion and  the plane of the sky (with $i > 0$ away from the observer). For the rest of the knots, only the tangential velocity, $v_t$, was obtained from the proper motion. 

\begin{figure}[htb]
\caption{\label{fig_knotsc_g3d}
Sketch of the 3D velocities of the post-collision knots C8, C10, C14, and C16 of the South outflow. The black arrows are velocities in the plane of the sky, derived from the proper motions. The blue arrows are the 3D velocities, with a radial component toward the observer.
}
\centering
\resizebox{0.8\hsize}{!}{\includegraphics{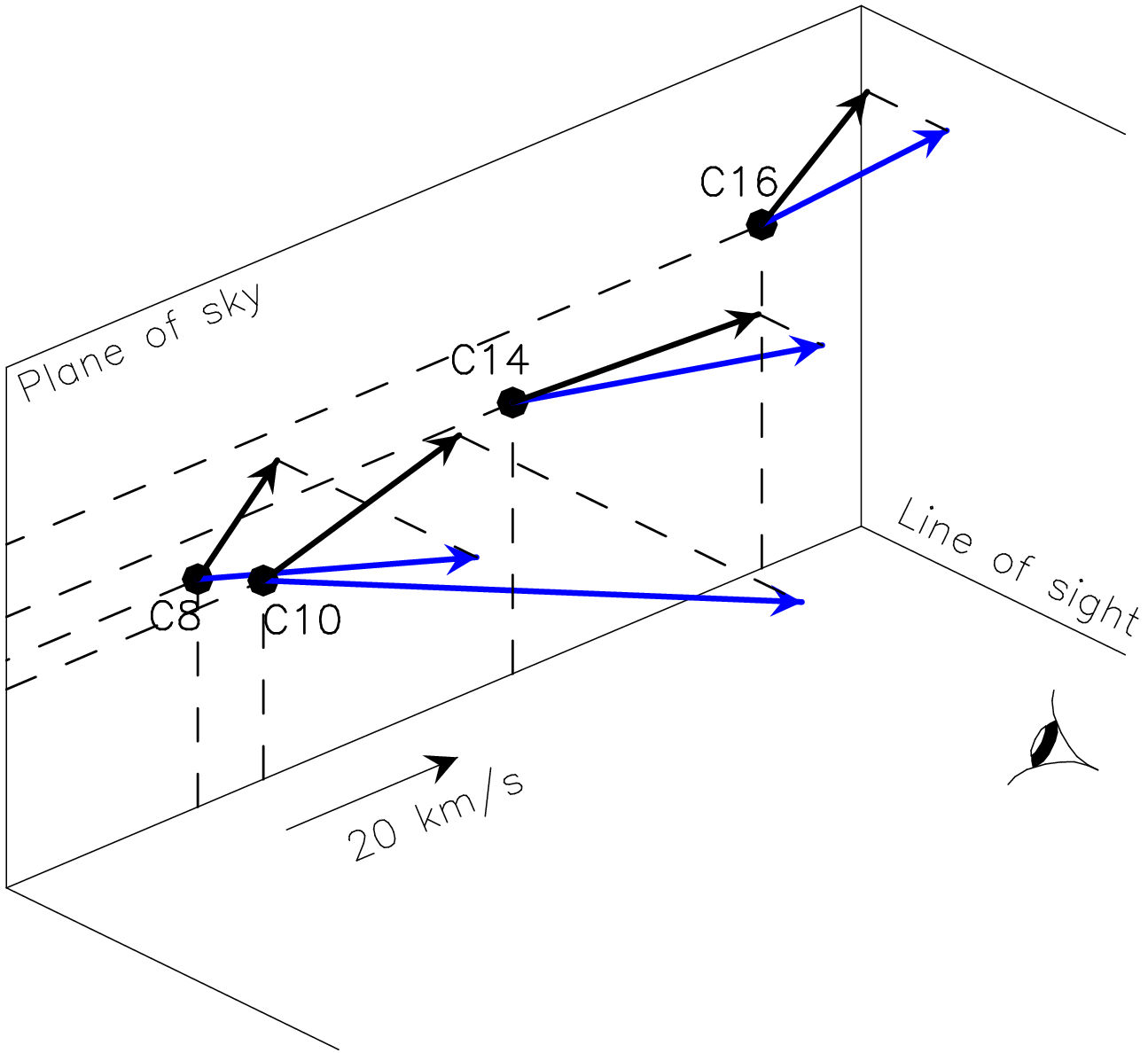}}
\end{figure}

\begin{figure}[htb]
\caption{\label{fig_knotsf_g3d}
Sketch of the 3D velocities of the pre-collision knots F1, F2-F3, F4, and F5 of East lobe of the South outflow. The black arrows are velocities in the plane of the sky, derived from the proper motions. The blue and red arrows are the 3D velocities, with a radial component toward and away from the observer, respectively.
}
\centering
\resizebox{0.8\hsize}{!}{\includegraphics{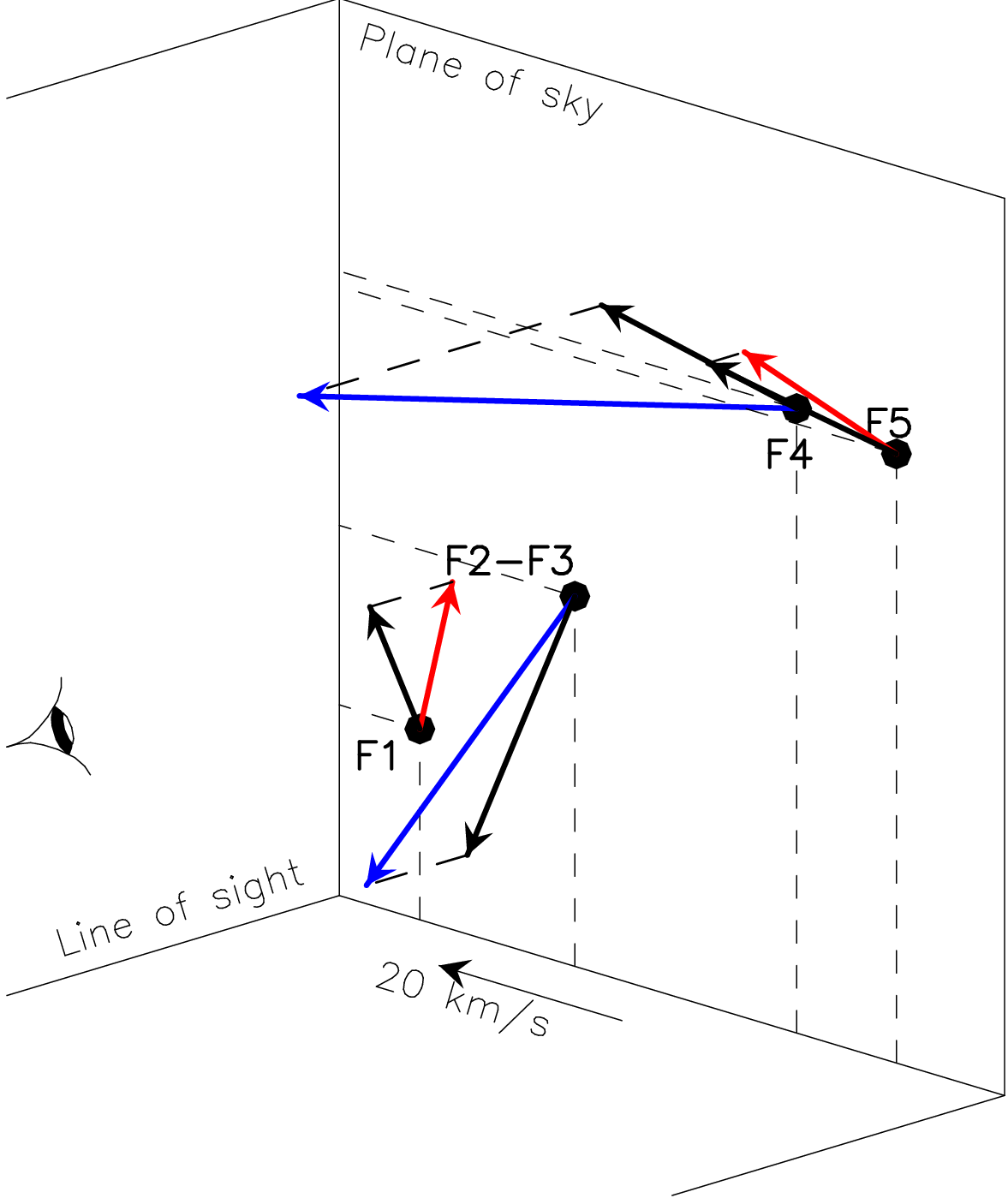}}
\end{figure}

Unfortunately the radial velocities of all the F, D, and C knots of the \hdos{} outflows could not be obtained because the long-slit positions did not encompass all their emission.
Knots C with observed radial velocity (C8, C10, C14, C16) have blueshifted velocities ranging from $\sim -10$ to $\sim -50$ \kms (see Fig.\ \ref{fig_knotsc_g3d}). 
Their radial velocity is fully compatible with belonging to the blueshifted lobe of the CO outflow. 
Regarding the knots F (F1, F2-3, F4, F5), 
CO(1--0) emission (see Fig.\ \ref{fig_co} indicates that knots F should be redshifted. 
Knots F1 and F5 are consistent with the CO data.
However, F2 and F4 are clearly blueshifted ($-10$ to $\sim -30$ \kms) (see Fig.\ \ref{fig_knotsf_g3d}).  
This may suggest that the knots F are tracing a bow shock with its axis near the plane of the sky, with knots F2-3 and F4 in the foreground, and knots F1 and F5 in the background \citep{Mas22}.

\section{Discussion}

The position and proper motions of the knots that are far from the interaction
area are consistent with 
D1, D2 belonging to the N-red lobe, and 
C1, C3, C4, C5 to the N-blue lobe of outflow N; and
F1, F2-3, F4, F5 belonging to the S-red lobe, and
T, C2 to the S-blue lobe of outflow S. 

The positions of these knots were used to derive the direction of the axes of the outflow lobes. 
Both outflow N and outflow S are slightly bent at the position of their respective exciting sources. 
The difference in PA of the red and blue lobes of outflow N is $9\fdeg8\pm0\fdeg6$. 
For outflow S, the difference in PA is $5\fdeg6\pm0\fdeg3$. 
For both outflows the bending at the position of the exciting source produces a concavity pointing roughly southward.

The intersection of the blue lobes of outflows N and S has been identified as the (possible) collision point, O. 
The area of (possible) interaction between the two outflows is westward of the collision point.
The position and proper motion of the knots in the area of interaction has made possible the identification of the knots of the blue lobes of both outflows (which we called post-collision knots) as belonging to outflow N (N-post), namely C6, C7, and C9; or belonging to outflow S (S-post), namely C8, and C10 to C17.

Both the N-post and the S-post knots are well aligned along an axis passing through the collision point, which is bent with respect to the corresponding blue lobe axis.
For outflow N, the axis of the post-collision knots deviates $+15\fdeg6\pm1\fdeg0$ (counterclockwise, northward) from its original pre-collision direction.
For outflow S, the deviation of the axis of the post-collision knots from its original pre-collision direction is $-8\fdeg0\pm0\fdeg3$ (clockwise, southward).
This is consistent with the collision scenario, with the N-post knots being pushed northward by outflow S coming from the south, and the S-post knots being pushed southward by outflow N coming from the north. 
The deviation angle is of the same order for both outflows, as would be expected from two outflows with a similar linear momentum \citep{Bel12}, confirmed by the similar value of the average tangential velocity of the pre-collision knots (Table \ref{tab_preaxis}).
However, the deviation angles are small, thus indicating that the interaction between the two outflows is most probably not a head-on collision, but a grazing collision.

The values of the linear momentum of the CO outflows reported by \citet{Bel12} are 0.06 \Msol~\kms for the blue lobe of outflow N and 0.09 \Msol~\kms for the blue lobe of outflow S.
The deviation of $15\fdeg6$ of the blue lobe of outflow N requires a change of its linear momentum of 0.017 \Msol~\kms, which is a 19\% of the linear momentum of the blue lobe of outflow S.
Similarly, the deviation of $8\fdeg0$ of the blue lobe of outflow S requires a change of linear momentum of 0.013  \Msol~\kms, a 21\% of the linear momentum of the blue lobe of outflow S.
These figures are consistent with a scenario of a collision of both outflows, with an interchange of the same amount of linear momentum, $\sim 0.015$ \Msol~\kms, which is a fraction of $\sim 20$\% of their pre-collision linear momentum.

The proper motion of the post-collision knots is consistent with their tangential velocities being radial with respect to the collision point, with an average difference of direction from radial $\Delta\mathrm{PA}=-8\degr\pm4\degr$,  but the rms dispersion is high, $\sigma_{\Delta\mathrm{PA}}=26\degr$.

The 3D velocity of a few knots was obtained using the line-of sight velocities obtained by \citet{Mas22}.
The radial velocity of knots C is fully compatible with belonging to the blueshifted lobe of the CO outflow. 
Regarding knots F, their 3D velocity suggests that they can be tracing a bow shock with its axis near the plane of the sky \citep{Mas22}.

\section{Conclusions}

There is strong evidence of interaction of the N and S outflows of IC 1396N, based on the following points.
\begin{itemize}
\item
There is a clear enhancement of shock-excited \hdos{} emission in the region of interaction.
\item
The deviations of the post-collision axes are consistent with an interchange of a similar amount of linear momentum between the two outflows.
\item
The collision point is located near the knots C5 and C6. 
Knot C5  is  the  knot  that  shows  the highest variability  between the three  epochs, and  knot C6 is the only knot with a significant change of morphology. 
They are most probably the closest knots to the collision point. 
\item
The motion of the post-collision knots is consistent with radial motion from the collision point.
\item
The small deviation of the post-collision knots implies that the interaction is not a head-on collision, but likely a grazing collision.
\end{itemize}

Additional observations would help to confirm the collision scenario.
In particular, a third epoch \hdos{} image, with a time interval of $\sim10$ years (that is, around 2027), would allow us to discriminate proper motions from changes in morphology of the knots.
In addition, for a good measurements of the radial velocities of more than a few knots, integral field spectroscopy of the region of interaction would be necessary.

\begin{acknowledgements}

We thank the referee for his/her helpful comments.
This work has been partially supported by the Spanish MINECO grants
AYA2014-57369-C3 and AYA2017-84390-C2 (cofunded with FEDER funds),
the PID2020-117710GB-I00 grant funded by MCIN/ AEI /10.13039/501100011033, 
and the MDM-2014-0369 of ICCUB (Unidad de Excelencia `Mar\'{\i}a de Maeztu'), and by the program Unidad de Exceencia `Mar\'{i}a de Maeztu' CEX2020-001058-M.

\end{acknowledgements}

\begin{appendix}

\section{Photometry of the knots and additional tables}

Table \ref{tab_phot} shows the differential photometry of the knots of the IC 1396N outflows of two previous images taken in 2003 \citep{Car06} and 2005 \citep{Bel09} with respect to the image obtained in the present work.
The rest of tables show the values for each knot used for computing the average values of
distances to the outflow axes (Tables \ref{tab_distpre} and \ref{tab_distpost}), 
radial PA from the collision point (Table \ref{tab_radpost}), and 
distances of the ballistic trajectories to the collision point (Table \ref{tab_postdist}).

\begin{table}[htb]
\centering
\caption{\label{tab_phot}
Differential photometry of the knots of IC 1396N:
difference in magnitude from the 2003 and 2005 images to the 2016 image of this work.
}
\begin{tabular}{lr@{$\,\pm\,$}cr@{$\,\pm\,$}c}
\hline\hline
     & \multicolumn{2}{c}{Epoch 2003\tablefootmark{a}} 
     & \multicolumn{2}{c}{Epoch 2005\tablefootmark{b}} \\
Knot & \multicolumn{2}{c}{(mag.)} & \multicolumn{2}{c}{(mag.)} \\
\hline
C1  & $-0.20$ & $0.11$ &  $+0.02$ & $0.08$ \\
C2  & $-0.11$ & $0.09$ &  $-0.02$ & $0.07$ \\
C3  & $-0.31$ & $0.07$ &  $+0.04$ & $0.04$ \\
C4  & $-0.18$ & $0.07$ &  $+0.07$ & $0.05$ \\
C5  & $-0.42$ & $0.03$ &  $-0.15$ & $0.02$ \\
C6  & $-0.08$ & $0.01$ &  $+0.05$ & $0.01$ \\
C7  & $-0.03$ & $0.01$ &  $+0.10$ & $0.01$ \\
C8  & $+0.07$ & $0.01$ &  $+0.14$ & $0.01$ \\
C9  & $-0.18$ & $0.03$ &  $-0.05$ & $0.03$ \\
C10 & $-0.21$ & $0.01$ &  $+0.05$ & $0.01$ \\
C11 & $-0.00$ & $0.01$ &  $-0.01$ & $0.01$ \\
C12 & $-0.04$ & $0.02$ &  $+0.03$ & $0.01$ \\
C13 & $-0.08$ & $0.01$ &  $-0.03$ & $0.01$ \\
C14 & $+0.04$ & $0.01$ &  $+0.06$ & $0.01$ \\
C15 & $+0.15$ & $0.02$ &  $-0.04$ & $0.02$ \\
C16 & $+0.02$ & $0.01$ &  $+0.04$ & $0.01$ \\
C17 & $+0.01$ & $0.03$ &  $-0.14$ & $0.03$ \\
D1  & $-0.01$ & $0.02$ &  $-0.01$ & $0.01$ \\
D2  & $-0.14$ & $0.02$ &  $-0.00$ & $0.02$ \\
F1  & $+0.01$ & $0.02$ &  $+0.21$ & $0.02$ \\
F2  & $+0.02$ & $0.02$ &  $+0.04$ & $0.02$ \\
F3  & $-0.32$ & $0.03$ &  $-0.00$ & $0.02$ \\
F4  & $-0.07$ & $0.05$ &  $+0.05$ & $0.04$ \\
F5  & $+0.02$ & $0.02$ &  $+0.09$ & $0.02$ \\
T   & $-0.36$ & $0.14$ &  $-0.06$ & $0.09$ \\
\hline
\end{tabular}
\tablefoot{ 
\tablefoottext{a}{From the image of \citet{Car06}}
\tablefoottext{a}{From the image of \citet{Bel09}}
}
\end{table}

\begin{table}[htb] 
\centering
\caption{\label{tab_distpre}
 Distance of the pre-collision knots to the outflow axes, passing through the exciting sources C (outflow N) and I (outflow S) (see Table \ref{tab_preaxis}).}
\begin{tabular}{lcr}
\hline
\hline
     &      & \multicolumn{1}{c}{$d$}  \\
Knot & Axis & \multicolumn{1}{c}{(arcsec)}\\
\hline
D1   & N-red  & $1.29$  \\
D2   & N-red  & $1.34$  \\
\hline
\multicolumn{2}{r}{$\langle d^2\rangle^{1/2}=$} & $1.32$ \\
\hline
C1	& N-blue   &	$0.96$  \\
C3	& N-blue   &	$0.52$  \\
C4	& N-blue   &	$0.02$  \\
C5	& N-blue   &	$0.03$  \\
\hline
\multicolumn{2}{r}{$\langle d^2\rangle^{1/2}=$} & $0.55$ \\
\hline
F1	  & S-red  & $0.98$  \\
F2-F3 & S-red  & $0.26$  \\
F4	  & S-red  & $1.38$  \\
F5	  & S-red  & $0.16$  \\
\hline
\multicolumn{2}{r}{$\langle d^2\rangle^{1/2}=$} & $0.86$ \\
\hline
T	 & S-blue  & $0.92$  \\
C2   & S-blue  & $0.11$  \\
\hline
\multicolumn{2}{r}{$\langle d^2\rangle^{1/2}=$} & $0.66$ \\
\hline
\end{tabular}
\end{table}

\begin{table}[htb]      
\centering
\caption{\label{tab_distpost}
Distance of the post-collision knots to the outflow post-collision axes.
passing through the collision point,
\RA{21}{40}{36}{71},
\DEC{+58}{18}{18}{1}, 
and with position angles 
$-106\fdeg7$ (outflow N) and
$-74\fdeg7$ (outflow S)
(see Table \ref{tab_postaxis}).}
\begin{tabular}{lcr}
\hline
\hline
     &      & \multicolumn{1}{c}{$d$}   \\
Knot & Axis & \multicolumn{1}{c}{(arcsec)} \\
\hline
C6   & N-post  & $1.09$ \\
C7   & N-post  & $1.52$ \\
C9   & N-post  & $0.81$ \\
\hline
\multicolumn{2}{r}{$\langle d^2\rangle^{1/2}=$} & $1.18$ \\
\hline
C8   & S-post & $1.61$  \\
C10  & S-post & $0.75$  \\
C11  & S-post & $1.74$  \\
C12  & S-post & $0.38$  \\
C13  & S-post & $3.85$  \\
C14  & S-post & $0.42$  \\
C15  & S-post & $1.02$  \\
C16  & S-post & $0.01$  \\
C17  & S-post & $0.45$  \\
\hline
\multicolumn{2}{r}{$\langle d^2\rangle^{1/2}=$} & $1.58$ \\
\hline
\end{tabular}
\end{table}

\begin{table}[htb]      
\centering
\caption{\label{tab_radpost}
Radial PA from the collision point to the post-collision knots position, 
and deviation from their tangential velocity position angle, $\Delta$PA.
The collision point considered is 
\RA{21}{40}{36}{71},
\DEC{+58}{18}{18}{1}
(see Table \ref{tab_rad}).
}
\begin{tabular}{lrr@{$\,\pm\,$}l}
\hline
\hline
      & Radial PA & \multicolumn{2}{c}{$\Delta$PA} \\
Knot  & (deg)     & \multicolumn{2}{c}{(deg)}      \\
\hline
C7	& $ -84$ & $-31$&$10$ \\
C9	& $-113$ & $180$&$43$ \\
C8	& $-117$ & $ 69$&$28$ \\
C10	& $ -70$ & $ -6$&$9 $ \\
C11	& $ -65$ & $-66$&$35$ \\
C12	& $ -77$ & $ 32$&$42$ \\
C13	& $ -94$ & $ 46$&$20$ \\
C14	& $ -73$ & $-20$&$5 $ \\
C15	& $ -71$ & $  1$&$25$ \\
C16	& $ -75$ & $ 19$&$10$ \\
C17	& $ -76$ & $  3$&$13$ \\
\hline
\multicolumn{2}{r}{$\langle\Delta\mathrm{PA}\rangle=$} & $-7.8$&$3.6$ \\
\multicolumn{2}{r}{$\sigma_{\Delta\mathrm{PA}}=$}      &\multicolumn{2}{l}{$26.4$} \\
\hline
\end{tabular}
\end{table}

\begin{table}[htb]
\centering
\caption{\label{tab_postdist}
Distances of the ballistic trajectories of the post-collision knots to the
collision point, 
\RA{21}{40}{36}{71},
\DEC{+58}{18}{18}{1}.
}
\begin{tabular}{lr}
\hline
\hline
      & \multicolumn{1}{c}{$d$}  \\
Knot  & \multicolumn{1}{c}{(arcsec)}   \\
\hline
C7    &  2.0 \\
C9    &  0.1 \\
C8    &  1.9 \\
C10   &  0.8 \\
C11   &  9.4 \\
C12   &  6.2 \\
C13   &  8.5 \\
C14   &  4.8 \\
C15   &  0.4 \\
C16   &  6.6 \\
C17   &  1.0 \\
\hline
$\langle d^2\rangle^{1/2}=$ & 5.0 \\
\hline
\end{tabular}
\end{table}
\end{appendix}
\end{document}